\PassOptionsToPackage{unicode,hypertexnames=false}{hyperref}
\PassOptionsToPackage{hyphens}{url}
\documentclass[
  english,
]{article}
\usepackage{xcolor}
\usepackage{amsmath,amssymb}
\usepackage{iftex}
\ifPDFTeX
  \usepackage[T1]{fontenc}
  \usepackage[utf8]{inputenc}
  \usepackage{textcomp} 
\else 
  \usepackage{unicode-math} 
  \defaultfontfeatures{Scale=MatchLowercase}
  \defaultfontfeatures[\rmfamily]{Ligatures=TeX,Scale=1}
\fi
\usepackage{lmodern}
\ifPDFTeX\else
\fi
\IfFileExists{upquote.sty}{\usepackage{upquote}}{}
\IfFileExists{microtype.sty}{
  \usepackage[]{microtype}
  \UseMicrotypeSet[protrusion]{basicmath} 
}{}
\makeatletter
\@ifundefined{KOMAClassName}{
  \IfFileExists{parskip.sty}{%
    \usepackage{parskip}
  }{
    \setlength{\parindent}{0pt}
    \setlength{\parskip}{6pt plus 2pt minus 1pt}}
}{
  \KOMAoptions{parskip=half}}
\makeatother
\usepackage{graphicx}
\makeatletter
\newsavebox\pandoc@box
\newcommand*\pandocbounded[1]{
  \sbox\pandoc@box{#1}%
  \Gscale@div\@tempa{\textheight}{\dimexpr\ht\pandoc@box+\dp\pandoc@box\relax}%
  \Gscale@div\@tempb{\linewidth}{\wd\pandoc@box}%
  \ifdim\@tempb\p@<\@tempa\p@\let\@tempa\@tempb\fi
  \ifdim\@tempa\p@<\p@\scalebox{\@tempa}{\usebox\pandoc@box}%
  \else\usebox{\pandoc@box}%
  \fi%
}
\def\fps@figure{htbp}
\makeatother
\ifLuaTeX
\usepackage[bidi=basic,shorthands=off]{babel}
\else
\usepackage[bidi=default,shorthands=off]{babel}
\fi
\ifLuaTeX
  \usepackage{selnolig} 
\fi
\providecommand{\tightlist}{%
  \setlength{\itemsep}{0pt}\setlength{\parskip}{0pt}}
\usepackage{hyperref}
\usepackage{float}
\IfFileExists{xurl.sty}{\usepackage{xurl}}{} 
\hypersetup{
  pdftitle={Auditable AI-Assisted Research Writing: An Engineering Discipline with Pre-Registered Process Observation},
  pdflang={en},
  hidelinks,
  pdfcreator={LaTeX via pandoc}}

\title{Auditable AI-Assisted Research Writing: An Engineering Discipline with Pre-Registered Process Observation}
\author{Yang Zhou\textsuperscript{a,*} and Chengqun Yu\textsuperscript{a}\\[4pt]
{\normalsize \textsuperscript{a}Institute of Geographic Sciences and Natural Resources Research,}\\
{\normalsize Chinese Academy of Sciences, Beijing 100101, China}\\[2pt]
{\normalsize *Lead contact and corresponding author: Yang Zhou (996420470@qq.com)}}
\date{}

\makeatletter
\long\def\@makecaption#1#2{%
  \vskip\abovecaptionskip
  \sbox\@tempboxa{#2}%
  \ifdim \wd\@tempboxa >\hsize
    #2\par
  \else
    \global \@minipagefalse
    \hb@xt@\hsize{\hfil\box\@tempboxa\hfil}%
  \fi
  \vskip\belowcaptionskip}
\makeatother
\begin{document}
\hypersetup{pageanchor=false}
\pagenumbering{gobble}
\maketitle

\subsection{Summary}\label{summary}

Language models now draft, classify and criticise inside research production, yet the artifacts they help produce carry little accountable history. Rather than detecting machine involvement afterwards, we specify an auditability discipline built at production time: git sealing with an anchor lineage, hash-bound provenance, red-line gates that refuse non-compliant artifacts and log every refusal, cross-model role separation, and programmatic assembly from registered sources. Adherence is instrumented by metric cards, each carrying a pre-registered blind spot and evidential standing, frozen before the prospective case it observes. In that case the observed project\textquotesingle s pre-registered confirmatory test was executed under seal and returned No-Go, and that project\textquotesingle s frozen stopping rule halted the work, against its own operators. A lower-graded retrospective case covers families whose machinery predates the protocol. Current observations are provisional; we release a package from which a third party can recompute every primary metric.

The Bigger Picture

A reader who wants to know where a sentence, a number, or a classification decision in a research paper came from has, at present, almost nowhere to look. Detecting machine involvement afterwards chases an adversarial surface. We build the record as work happens: version-control sealing, hash-bound provenance, refusal-logging gates, cross-model role separation, and scripted assembly from registered sources, under a protocol frozen before the prospective case it observes; a retrospective case is reported apart, at lower standing. In the prospective case the observed project\textquotesingle s pre-registered test returned No-Go, and that project\textquotesingle s frozen stopping rule, not ours, halted the work, against its own operators. We claim nothing about whether this improves the research it governs, only what the mechanisms recorded. Should such records become ordinary, provenance becomes something readers recompute.

\textbf{Keywords:} research provenance; auditability; AI-assisted research; pre-registration; reproducible research workflows

\subsection{Introduction}\label{introduction}

Partway through the production run this paper observes --- a cross-year lineage study of provincial funding-call guidelines --- that project\textquotesingle s pre-registered confirmatory test was executed under seal against endpoint years its operators had never read, and the no-go check at its centre evaluated to TRUE. The frozen rule\textquotesingle s response to that boolean was not a footnote but a halt: the confirmatory claim that test was to support was withdrawn rather than reinterpreted, no quantity recomputed under the triggered revision path enters this round of reporting, and the project returned to protocol revision under a new pre-registration cycle. We open there because that outcome is the best evidence we can offer for what this paper is about: not a finding but a discipline, and the test of a discipline is what it does to the people who wrote it at the moment it costs them something.

Language models now occupy the load-bearing middle of research production. They screen literature, classify source material, propose and criticize analyses, and draft prose. The output is frequently good. What it is not, by default, is accountable. A manuscript is a compressed statement of conclusions, nearly silent about provenance, and it has always been so. A non-deterministic collaborator does not create the gap between an artifact and its history, but widens it quickly enough that the usual tolerance stops being tenable.

Two responses are available. The first is detection: classifiers that estimate whether text was machine-generated, stylometric audits, watermarking schemes. Detection is post hoc forensics on an adversarial surface, and adversarial surfaces do not converge: every improvement in detection is an input to the next round of evasion. The second is auditability: build the record at production time, so that "where did this sentence, this number, this classification decision come from" becomes a question a machine can answer from artifacts that already exist. This paper takes the second route, as an engineering problem rather than a normative appeal.

Auditability has been urged for a long time without being adopted. Result-blind review was proposed as early as the 1960s {[}1{]}; journal editors made prospective registration of clinical trials a condition of publication, then restated what counts as registered {[}2{]}, {[}3{]}; statcheck demonstrated that a purely mechanical consistency check over published statistics finds errors at scale {[}4{]}; nanopublications and citation-context services showed that machine-readable claim units are constructible {[}5{]}--{[}8{]}. In these cases technical feasibility preceded broad adoption, often by years or decades. Our reading of that pattern, offered as a design hypothesis rather than as a measured result, is that the binding constraint was less the technology than the marginal cost an individual researcher had to pay, unfunded and unrewarded, to record and curate process. That is the cost machine assistance plausibly changes. The same machinery that drafts text can write manifests, maintain ledgers, bind claims to evidence rows, re-derive artifacts from their sources, and refuse to continue when a check fails. Recording and curation, historically the expensive part, is --- we argue, and do not measure --- much of what that machinery now performs at low marginal cost, which converts auditability from an exhortation into a specification question: which mechanisms, wired how, measured by what.

Our first contribution is that specification. We describe five mechanisms operating as one workflow: git sealing with an explicit anchor lineage, so every reported state corresponds to a named and verifiable point in history; hash-based provenance that binds each quantitative statement to a true source rather than to a retyped copy; red-line gates that refuse non-compliant artifacts and log every refusal, including refusals directed at the project\textquotesingle s own operators; multi-agent role separation with cross-model adversarial review, so that the party that produced a given artifact is never, for that artifact, the party that clears it; and programmatic body injection, under which no reported number is typed by hand and the manuscript is assembled from sources by script. None of these is new to software engineering. The contribution is their transposition into research writing as an end-to-end discipline a group can adopt, and their instrumentation, so that adherence becomes measurable rather than asserted. What the present round delivers against that instrumentation is a first and deliberately unflattering set of readings: many cards resolve to an explicit missing code or an undefined value, and Section 6 reports that coverage rather than a completed measurement of end-to-end adherence.

The second contribution is how we know. Before the observed production began, we froze an observation protocol fixing what would be measured, how, and against what criteria: twenty-one metric cards across seven families, each specifying its definition, data source, evidential status and --- the field that matters most for honesty --- its own measurement blind spot. Missing values are first-class, with a dedicated four-value encoding, and missingness is itself a reported quantity. Aggregation across cases and families is prohibited by the protocol, not discouraged by taste. Because the protocol predates the confirmatory production it observes, that case\textquotesingle s evidence is prospective, which separates it from retrospective case retellings, however detailed.

Three commitments recur throughout. The first is prospectivity: the instrument was frozen first, and all activity preceding that freeze is labelled exploratory and excluded from confirmatory reporting. The second is reflexivity: the instrument obeys its own rules --- protocol revisions pass through an amendment ledger, its schema validator has refused ledger entries submitted by the project\textquotesingle s own orchestrating role, and the freeze ceremony is itself logged. The third is the dignity of negative results: the pre-registered confirmatory test that fired was executed as written, and reporting it in full is the pressure test of everything else described here.

The boundaries are stated at the outset and revisited in Section 7. There is no control group, so we make claims about system operation only, and none about the quality or efficiency of the research the workflow produces. We are simultaneously the observers and the observed, in a single laboratory, across one prospective and one retrospective case; no arrangement of this paper removes that conflict, and Section 7 describes how it is bounded rather than implying it away. Adjacent systems --- in-editor process archives, git-stored reasoning trees, agent metadata provenance, signed logs with autonomy grading --- address parts of the same problem, and Section 8 places our work against each. Section 2 presents the discipline and Section 3 the protocol; Sections 4 and 5 the prospective and retrospective cases; Section 6 reports the measurements, card family by card family, each beside the blind spot that limits it; and Section 9 states what is released, on what terms, and what is withheld.

\subsection{The Discipline}\label{the-discipline}

We call this a discipline rather than a tool: none of its parts is software a group installs; each is a rule about what must exist, in a recomputable form, before a step counts. The five mechanisms below sit on four registration cards in our observation protocol rather than one card each, and the mapping is worth stating once so that later sections can be read against it: git sealing is registered as M2; hash provenance, canonical-source binding and the claim--evidence index together as M1; the red-line gate as M3; and the sanitized external workspace through which cross-model review runs as M4. Each card carries a provenance record, the quantities it makes observable, and its own measurement blind spots, registered in advance rather than discovered afterwards. Two things below have no card of their own. Role separation itself is a governance control, not a metric. Programmatic body injection is assigned to no single card by deliberate protocol note, because the instruments that observe it --- the build-chain family and the gate-attempt family --- straddle M1 and M2.

The first mechanism is git sealing and anchor lineage. Every change to an engineering artifact lands as a commit, so no document is quietly edited in the working tree until publication. The three freeze-stage terms --- protocol, calibration, release --- may not stand in for one another, and a freeze is not a heading that says frozen: it is a manifest registering a path and a SHA-256 digest for every protocol document, canonical source, script, schema, ledger and environment lock, together with the commit, the tag and the recorded preflight exit codes, recomputed item by item by a verifier for which a missing commit reference is a default failure. Thereafter changes move only through an append-only amendment ledger, and failures only through a deviation ledger.

The second is hash provenance and canonical-source binding. Every derived artifact descends from one canonical source of truth, and every evidence entry terminates in a source identified by digest or by a versioned locator, rather than by citation string. A source is valid in exactly two ways: a local object whose recomputed digest matches its registration, or an external reference anchored by locator and date. Sources pending or failing verification are never folded into the valid set and are reported separately, since a blended rate hides exactly the cases a reader needs. On the measurement side, the observation protocol of Section 3 adds a condition of its own: the three read paths --- summary, derived audit view and authoritative registry --- must agree entry by entry at the same commit, and disagreement invalidates the observation snapshot rather than licensing a choice.

The third is the red-line gate: a non-optional automated check run before an artifact crosses an engineering boundary. Two rule families, one for protected relational information, one for scope-of-disclosure language, run at two scopes --- internal and public, the latter stricter --- returning pass, block, or usage error. The rule carriers enter neither version control nor the manifest\textquotesingle s hashed inventory, since a digest over a word list is itself a dictionary-attack oracle. Hits surface only as opaque codes and counts: the checker reports that something matched, not what. When a person overturns a block, a ledger row recording time, task, artifact count, verdict, reason and releasing role is mandatory; an override without one counts as never having happened. But a gate stops only what its rules recognize; false negatives are unmeasurable, so we claim no absence of leakage, only that every block and override is recomputable.

The fourth governs division of labour across agents and cross-model review. Roles are frozen before observation: an operator produces inside the observed repository; an observer reads and computes but never writes to it; an adjudicator settles disputes and may not adjudicate a sample they produced. Cross-model review runs through a sanitized external workspace: external models execute only outside the engineering root, only named files are copied in and only after clearing the gate at public scope, a placeholder specimen replaces the real artifact wherever a task needs structure rather than substance, any failure aborts and purges the workspace rather than exporting part of it, returning artifacts clear the same gate before re-entry, and each preparation and collection appends a ledger row. Anything pasted directly into an external model bypasses this channel and is unobservable; and in a single-laboratory effort one seat holds both operator and adjudicator identities, a conflict we register as structural rather than resolved.

The fifth is programmatic body injection. Released body text is never hand-typed: source markdown and a registered build script deterministically produce the artifact, and a read-only runner replays that build in a clean environment from the manifest\textquotesingle s script digest, not the working tree\textquotesingle s, then compares the result byte for byte with the repository artifact. On the measurement side the protocol adds that a replay which errors is coded INVALID and may never be recoded as no difference found. Bound to it is the claim--evidence index: every outward claim becomes an addressable atomic claim with at least one evidence entry descending through location, formula, parameters, hashed source and quality control, generated from one source of truth. The index has its own publication gate, which recomputes a frozen invariant set over the manuscript and its indices before granting release; it is a distinct gate from the red-line check described above, which governs disclosure rather than closure. The denominator is not drawn by the party being measured: a deterministic parser, under criteria frozen in advance, marks which sentences ought to have been registered, so an unregistered claim is machine-discoverable. Those criteria can still miss difficult claims, so candidate-sentence recall must be examined separately through human audit, with its sample size and adjudicator-overlap limitations reported alongside.

These five form a discipline rather than a checklist because they close on one another (Figure 1): the gate\textquotesingle s outputs are sealed by the manifest, the manifest is what the read-only runner replays from, and the external workspace cannot move a file without the gate. One-way read-only access, the stricter export scope, keeping rule carriers out of version control and recording only institutional roles as signatories are governance controls, not metrics: their observance appears in the deviation ledger and is not converted into a rate, because a discipline that scores itself on its own housekeeping has begun to measure the wrong thing.

\begin{figure}[H]
\centering
\pandocbounded{\includegraphics[keepaspectratio,alt={Figure 1: The five mechanisms close on one another: the gate\textquotesingle s outputs are sealed by the manifest, the manifest is what the read-only runner replays from, and the external workspace cannot move a file without the gate.}]{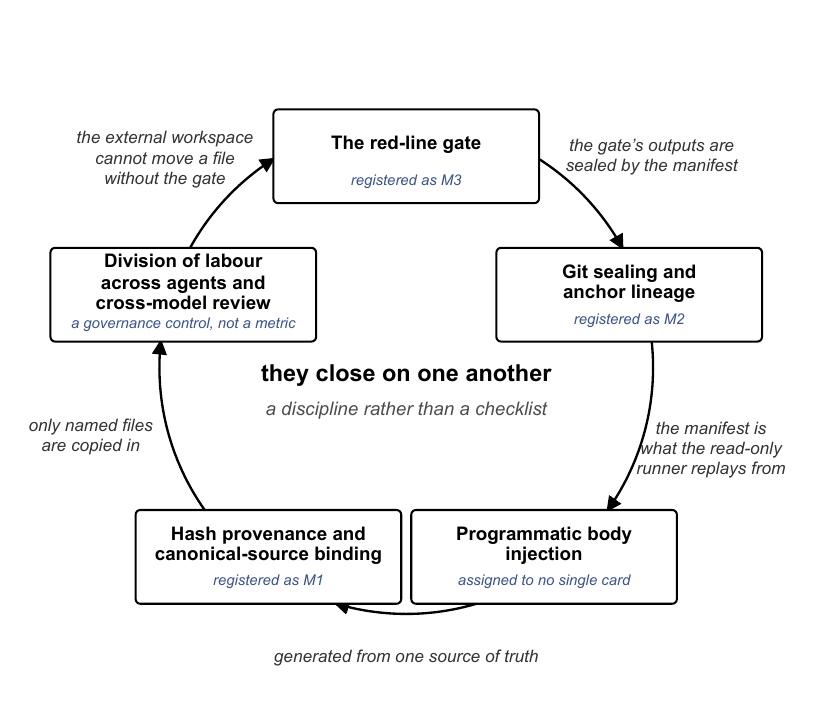}}
\caption{Figure 1: The five mechanisms close on one another: the gate\textquotesingle s outputs are sealed by the manifest, the manifest is what the read-only runner replays from, and the external workspace cannot move a file without the gate.}
\end{figure}

\subsection{Pre-Registered Observation Protocol}\label{pre-registered-observation-protocol}

The instrument was frozen before the confirmatory production it observes: the audit-package pre-registration protocol fixes what is measured, how each quantity is computed and what counts as a decision criterion, and is itself sealed under the discipline it measures. That ordering is what prospectivity means here, since a retrospective report can always choose afterwards which quantities flatter the workflow. The protocol therefore carries no status of its own, because sealed text that reports its own state is a second source of truth that expires silently; whether it is frozen and which anchor is current live only in the current freeze manifest and its strict verification output.

The measurement surface is 21 metric cards in seven families: build-chain integrity, claim-evidence closure, gate efficacy, sealing, privacy and caliber gating, human-machine attribution, and reproducibility and cost. Caliber, throughout this paper, denotes the granularity and rule set under which a quantity is computed; the same measure read at two calibers may differ without either reading being wrong. Every card follows one 14-field template, running from observed entity and event record through numerator, denominator, de-duplication unit and missing-value coding to adjudicating role, computation script, data source, standing, anti-gaming rule and measurement blind spot. Standing is assigned before any data exists, in four tiers: confirmatory, secondary, descriptive, exploratory. The anti-gaming field states in advance how that measure could be inflated -\/- by splitting commits or model calls, or by manufacturing gate hits -\/- and several cards were demoted to descriptive or exploratory on that ground before collection. The blind-spot field is transcribed verbatim wherever the card is reported. A standing clause holds that any value a release gate forces to a given level is a consistency check, not evidence of efficacy: the observable is the attempt process, and every attempt, including failures, is kept append-only.

Missing data is a first-class citizen. Four codes are frozen: MISSING for what should have been collected and was not, NA for the structurally inapplicable, PENDING for the in-flight, INVALID for what fails an equivalence check. None may be silently dropped or replaced by zero or one, and missing counts are themselves reported quantities. A zero denominator yields a null reported as UNDEFINED -\/- never 1.0, never a percentage -\/- and may not be composed downstream as though it were one. Both rules must be restated on every card; a card that merely inherits them is non-conforming, and a linter parsing all 21 cards against the field contract blocks manifest generation when one fails. Whether a data source exists is settled by the collection script at run time, not by protocol text, and a card whose source proves absent is reported under its pre-committed standing with its missing code. The reporting plan is fixed with the same specificity: milestone trajectories, no composition across the two cases, and no statistical inference beyond one interval method chosen at freeze time.

The protocol also takes its own medicine. Amendments move through three tiers: typographic fixes and clarifications apply globally, the latter obliged to argue in the record that no computed result changes, while substantive changes -\/- to a definition, a numerator, an inclusion rule, the case roster or a script\textquotesingle s logic -\/- take effect for future events only and are barred altogether once the window closes. Every amendment carries a blindness declaration recording whether the proposer had already seen the relevant results; where they had, dual-caliber parallel results are mandatory and the old results retained. A deviation ledger of the instrument\textquotesingle s own, kept in a namespace distinct from the ledgers of the projects it observes, has eight mandatory triggers -\/- among them every gate attempt whose registered verdict is FAIL or INVALID, every degraded data path, every manual override and every case in which role separation was breached rather than held -\/- and the attempt schema refuses a failing entry that references no deviation. That list binds this instrument; each observed project registers deviations under its own frozen rules and its own identifiers. Neither ledger copies its schema -\/- both reference the upstream schema by path, anchored in a sidecar and in the manifest -\/- so drift fails the validator at once.

Self-application shows most clearly in the protocol\textquotesingle s repeated refusal to certify itself. One freeze ceremony failed strict verification because a sealed run product embedded its own timestamps, so working-tree bytes, committed blob and registered hash could not agree; the answer was a rule rather than a patch -\/- sealed objects classified as static, regenerated or immutable record, volatile metadata moved to a sidecar outside the seal, and a repeated byte-identity check made a precondition of manifest generation. Later candidates failed differently: one had a live-ledger entry rewritten in place, and on another a successful attempt sat at an unanchored chain tail where altering one byte still left both audits green, which produced a tool-written checkpoint, an annotated tag as chain terminus and a fault-injection test aimed at that byte. A third class recurred -\/- a check enforced before the object it checks can exist -\/- closed by making such checks advisory in the ceremony window and mandatory under audit, with a timing audit that blocks manifest generation when the ordering is violated. Each episode carries its own deviation identifier; the failed anchors were left in place.

The same discipline decides when the confirmatory observation window may open. Its start is a single non-retroactive instant fixed by rule -\/- no earlier than the latest of three independently provable events -\/- and adjudicated by independent review on machine-checkable evidence, the protocol writing the rule and never the value. The fifth dual-repository review returned NOT\_ADJUDICATED: one bounding event could not be proved from the objects themselves, a lightweight tag persisting no creation time of its own, and approximation in its place is barred. An instrument that will not date itself until it can prove the date is one whose freeze is load-bearing. Only at the seventh review, after two further rounds of repair, with each bounding event readable from an annotated tag object and both audit calibers re-run by the reviewers, was a single effective\_at adjudicated and the window opened. Everything collected beforehand keeps pilot or exploratory status permanently and may not enter any confirmatory numerator, denominator or mechanism claim; the adjudication opens the window forward and licenses no backdating.

Two objects share the word confirmatory in this paper and are kept apart throughout. A confirmatory test is a pre-registered decision procedure belonging to an observed project, evaluated once against a rule fixed in advance; its outcome is a boolean, and Section 4 reports one. Confirmatory standing is the evidential grade this protocol assigns to its own measurements, governed by the window above and by the snapshot-status rules under which each collection run is classified. The two move independently: a project can execute its confirmatory test while the snapshot recording that event still lacks confirmatory standing, and the outcome of such a test neither raises nor lowers the grade of any card. What the instrument cannot do is stand outside the team that built it, a limitation stated in full in the boundaries section.

\subsection{Case 1 (Prospective): A Policy-Corpus Engineering Project}\label{case-1-prospective-a-policy-corpus-engineering-project}

CASE-01 is a policy-corpus research project --- a cross-year lineage study of provincial science-and-technology programme guidelines for two jurisdictions --- held in its own repository and registered in the frozen case roster as this work\textquotesingle s single confirmatory case. Its role is prospective and instrumented: the observation protocol was frozen before the project entered its confirmatory phase, and the window opens at a single \texttt{effective\_at} fixed by rule rather than by choice. The rule requires that start point to fall no earlier than the latest of three source-verifiable events --- the tagger times of the project\textquotesingle s current protocol-freeze tag and of the observation repository\textquotesingle s current checkpoint tag, and the later filesystem timestamp of two strict-verification evidence files --- and it forbids earlier activity from entering any confirmatory numerator, denominator, or mechanism claim, or being absorbed retroactively. Everything before that point is permanently labelled pilot/exploratory. The team that operates the project and the team that operates the instrument are the same people; we treat that as the governing threat to this case and return to it in the limitations section.

Freezing was itself contested. Successive rounds of independent review rejected candidate anchors in both repositories before one passed strict verification; every rejected candidate was left in place with its failure recorded, not deleted and re-cut. An anchor is an annotated tag whose manifest binds hashes for the protocol text, the tooling, and the canonical source files; a separate seal manifest binds each required artifact role to the anchor commit under an explicit state machine, so that which files were sealed, in which role, and under which commit is machine-checkable. When two frozen rules collided in production --- a scan designed to keep result values out of dispatched task packages structurally rejecting a package whose payload was the machine-generated evidence itself --- we escalated rather than patched around it. The adjudicated response became a standing formula: amend the criterion, re-run the full regression suite, obtain one external review round, and re-mint the anchor completely. The formula was applied again, without re-escalation, when the same class of collision recurred. Each application produced an amendment entry and a deviation entry; each superseded anchor was declared void, while artifacts produced under it retained their identity through repository history and the anchored verification recorded at the time.

Review was cross-model by construction: production sessions ran on one model, independent review on models from other vendors at fixed reasoning settings, with a further internal review pass. Two external rounds terminated at the vendor\textquotesingle s content filter before an opinion could be written; only after the task order was rewritten in defect-report form --- input construction, expected exit code, observed exit code, source location --- did the external reviewer complete and deliver. We record this as a fact about the review channel, not about the code under review. The substantive finding was a coverage gap: the external reviews returned PASS over a small set of constructed inputs, while the internal review, with a far larger set, produced findings a sample of which were re-run individually, every one confirmed; the remainder were recorded as reported defects not yet adversarially re-run and were explicitly barred from being counted as confirmed. The team declined to treat the external PASS as a release basis; release followed a separate adjudication after repair and extension of the regression suite. Gate attempts were retained append-only throughout: the case ledger holds 24 attempts, of which 6 recorded at least one gap; the freeze-ceremony gate did not pass on its first attempt; and the recurrence card separately records 2 closed gap segments and 3 later attempts that failed at a gate after that gate had earlier passed. Those two counts have different units and are not composed; 2 gate-level failure segments remained open at the current snapshot.

Three governance events are worth reporting because in each the gates acted against the operators, not the model. First, the standing formula above converted a collision between frozen rules from an occasion for discretion into a mechanical procedure. Second, a task-completion notice arrived in non-standard form carrying embedded instructions; it was refused as an instruction source, and verification proceeded only from machine-readable artifacts on disk; the standing rule that followed is that completion is established by landed artifacts, never by a notice. Third, the gates blocked the operators repeatedly. A schema validator rejected the orchestrating role\textquotesingle s own ledger entries. An access log required registration before reading. An isolation scan rejected dispatched packages. And two scripts lacking a command-line parser executed their main flow when probed with a help flag and overwrote an out-of-repository key, rendering a sealed ciphertext permanently undecryptable; the unsealing path was re-derived from the committed plaintext hash list, and a rule was added that repository tools must be read before invocation.

The ledger accounting for these events belongs to the observed project\textquotesingle s own deviation ledger rather than to the instrument\textquotesingle s. Each application of the formula sits at DEV-G1-012 and DEV-G1-013, the isolation scan\textquotesingle s rejection of a dispatched package at DEV-G1-011, the overwritten key at DEV-G1-005; those identifiers are governed by that project\textquotesingle s frozen rules, not by the eight-trigger list of Section 3, which binds the instrument\textquotesingle s ledger only. The completion notice and the refusal of the orchestrating role\textquotesingle s own entries carry no deviation entry, and correctly so: each is a gate enforced successfully rather than a gate that failed --- no attempt was registered FAIL or INVALID, and role separation was held rather than breached --- so neither falls under a mandatory trigger of either rule set. Both are recorded instead in the project\textquotesingle s breakpoint archive, the dated continuity document in which that project carries its own narrative record between working sessions. In none of these events was the difficulty resolved by amending the record.

The endpoint stage ended in a fired gate (Figure 2). Its final step was one irreversible evaluation: a verifier-only entry point, given the anchor commit and the seal manifest, computed the pre-registered decision rule NG-H1 against a blind baseline the operator never read, printed the boolean, exited non-zero, and wrote a decision artifact committed with its hash. NG-H1 returned TRUE. Under the frozen rule this means the pre-registered confirmatory test was evaluable and returned No-Go; it is not the NOT\_EVALUABLE outcome, for which a contingency plan had been written and sealed in advance. The two senses separated in Section 3 both apply and must not be merged: the confirmatory test was executed under seal and returned No-Go, so the confirmatory claim it was to support is withdrawn, while the observation snapshot that records the event carries provisional status and pilot/exploratory data class, so the halt enters this paper as a process fact rather than as a confirmatory measurement. The team recorded that the sealed plan\textquotesingle s terminal-class template does not directly apply and that fresh adjudication is required, rather than stretching the template to fit. Halting was itself specified: no advance to the pass-criteria adjudication, no new production sessions, no rewriting of the paper\textquotesingle s claim tier, no edit to the lineage protocol text, and the anchor and its seal state left untouched. The endpoint-side input displays remain under a standing interpretation ban, and the boolean is the only outcome reported here. What this case supports is a system claim --- an instrument frozen before the production it observed, a decision rule evaluated mechanically, and a negative verdict that propagated into a halt with no discretionary step. It supports no claim that the discipline improved the quality or the speed of the research it governed.

\begin{figure}[H]
\centering
\pandocbounded{\includegraphics[keepaspectratio,alt={Figure 2: The endpoint stage: from the sealed anchor and dispatched task packages to the verifier-only evaluation of NG-H1 and the halt its boolean triggered.}]{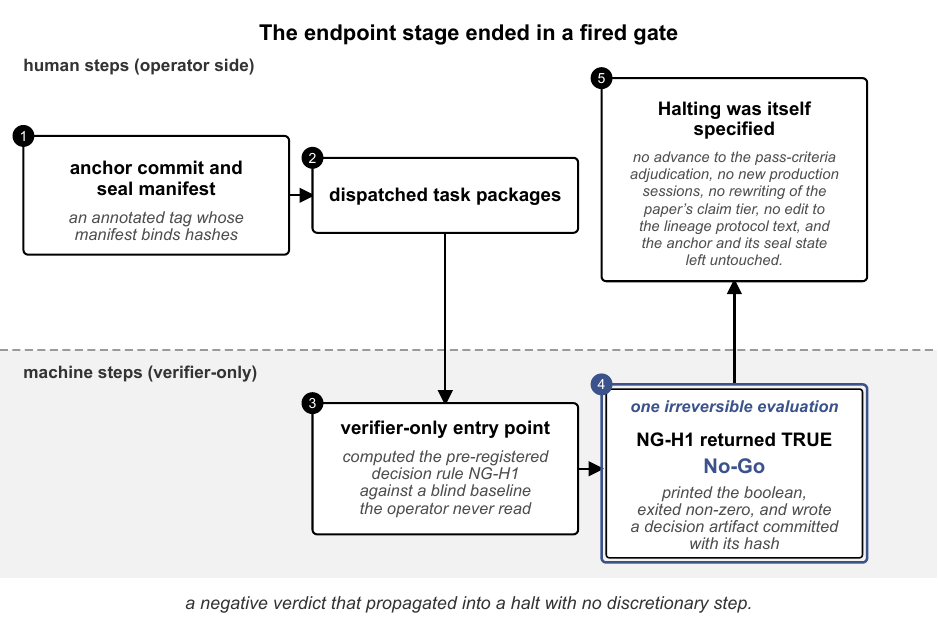}}
\caption{Figure 2: The endpoint stage: from the sealed anchor and dispatched task packages to the verifier-only evaluation of NG-H1 and the halt its boolean triggered.}
\end{figure}

\subsection{Case 2: Retrospective Corroboration}\label{case-2-retrospective-corroboration}

Case 2 is a completed conference-presentation engineering project: a talk, its slide deck, verbatim script, and contingency material, produced for an already-accepted paper. It enters this paper for one reason and supports only what that licenses. It is the sole instance in which the discipline of Section 2 had already been run end to end before the observation protocol existed, and thus the only historical source for the privacy-and-caliber gate and human-attribution families. Two of the four registered mechanism cards --- the red-line gate and the sanitized external workspace --- have their first working implementations in this project\textquotesingle s tool chain, so that part of the mechanism lineage is inspectable at its origin rather than merely narrated.

We state the evidence-grade difference before any figure. Case 2 is retrospective: the work had finished, its sealing chain and ledgers already existed, and the observation window was drawn afterwards. Its records were produced under no pre-registration constraint, which admits a systematic selective-recording risk that no later analysis removes. The frozen case registry fixes the arithmetic: the confirmatory case count is one, Case 1 alone; the reporting count is two, stratified. Prospective and retrospective results are never composed into a cross-case rate; statements spanning the two cases are qualitative, and every Case 2 figure carries a retrospective label.

Instrument coverage is partial, and the pattern of absence is the informative part: the families missing from Case 2 are exactly those whose machinery the discipline acquired afterwards. It supports no build-chain card, because neither a rendition manifest nor a body-integrity checker existed; no claim--evidence-closure card, because there was no claim index; no gate-efficacy card, because gate attempts were never recorded in structured form; and not the independent-rebuild card, because no drill was run. What it does support is the gate-activity, override-ledger-completeness and false-positive-adjudication cards, the human-decision-point and override-event cards, the exploratory multi-agent-activity card, and the cold-start archive card; the sealing-density and cost cards are registered for it at restricted availability. Neither list is exhaustive.

Four limitations travel with every Case 2 number. Commit granularity and milestone-marking conventions differ from Case 1, so Case 2\textquotesingle s sealing density stays in the ledger as a raw count, enters no card table in Section 6 and is never placed beside Case 1\textquotesingle s. Milestones are marked by breakpoint-archive documents rather than a systematic tagging protocol, which coarsens the time anchors. The external-call ledger is maintained by hand, and rows whose timestamp is a date plus a review note rather than an exact instant are force-coded as imprecise (8 such rows). The false-positive adjudication card is coded MISSING here, because no adjudicator verdict was registered in the required form, and an unadjudicated machine hit may not be counted on either side. Cold-start archives exist, but an archive is not a drill; no recovery drill was performed, and the protocol requires this be registered as not attempted rather than vanish as an empty denominator.

What the surviving records document, retrospectively and under the partial instrumentation set out above, is the following. Across 55 gate runs covering 163 scanned files, 12 runs were blocked; 8 of 8 human-override events carry all six ledger elements the protocol requires; the sanitized external workspace mediated 40 task packages, with 1 aborted and purged after a failed input scan. The mechanisms ran, leaving machine-readable traces a third party can recount, within the claim boundary of Section 7.

\subsection{Results}\label{results}

The frozen analysis plan fixes the reporting order: process quantities first, terminal values after, under the standing clause set out in Section 3. Gate attempts therefore open this section, and the build-chain and claim-closure families follow. One distinction governs how the rest should be read: only the gate-efficacy family is admitted as evidence about mechanism efficacy, and every other family reports a consistency check, a descriptive count or an exploratory quantity, so each family\textquotesingle s numbers carry the weight its pre-registered standing assigns them and no more. Every quantity below is injected programmatically from one collection run, none transcribed by hand. That run records the digest of each upstream canonical file before and after collection, and the snapshot\textquotesingle s read-only proof carries both sets. All card observations carry status PROVISIONAL, capped there because the observation freeze manifest is not yet in effect, with data class pilot/exploratory; no observation in this section carries confirmatory standing. The prospective and the retrospective case stand side by side and are never composed into a rate. Four cards are registered in the frozen set as spanning both cases; their raw counts are reported as registered, and no cross-case rate is derived from them. Each family is reported together with the measurement blind spot frozen on each of its cards, transcribed card by card rather than merged into one sentence per family; the English is a faithful rendering of the frozen source field, with nothing added, combined or dropped.

Gate efficacy is the only family admitted as mechanism evidence, measured on attempts, not outcomes; append-only retention of attempts is recorded on the card as Yes. First-pass performance (C1) is 1 of 2 gates. Gaps (C2) are split by gate and never summed across gates, 8 at the regression gate and 9 at the freeze ceremony, with per-attempt median 0.0 and maximum 6. Remediation (C3) is tracked as gate-level failure segments: 2 closed and 2 still open. Defect recurrence (C4) is reported rather than dropped for being unflattering, as two counts on two units, never composed: 2 closed gap segments, and 3 attempts that failed at a gate after that gate had earlier passed; the card carries no observation-window field, so the window length its blind spot requires alongside is not available at this snapshot. A caliber degradation is declared, not hidden: the upstream ledger has no milestone field, so the pre-registered milestone-by-gate unit degrades to gate level, that dimension recorded MISSING. Blind spots as frozen, card by card. C1: "If the upstream attempt ledger is not append-only (fixed output path, preflight sub-items masked by a summary bit), the process quantities of the C family cannot be recomputed. Adjudication rule: at collection time, if the upstream attempts ledger lacks append\_only=true or lacks a per-attempt immutable artifact directory, that batch of C-family data is uniformly marked pilot and a deviation is registered. For the historical adjudication see §14." C2: "Rule granularity differs across gates and is not comparable; summing across gates is forbidden." C3: "Wall-clock duration mixes in sleep, waiting and scheduling and does not equal work volume; continuity across time zones and across sessions cannot be discerned." C4: "If the observation window is short, recurrence has no way to show itself; the window length must be reported alongside." Blind-spot fields are transcribed as frozen, internal cross-references included: C1\textquotesingle s pointer to §14 is to a section of the observation protocol document, not of this paper.

Build-chain integrity asks whether a body artifact replays byte-for-byte from its registered script and sources (A1), whether five registration elements recompute equal (A2), and what share of source characters carries an authorship record (A3). A1 and A2 both resolve to denominator 0: no body rendition of the observed manuscript had been registered upstream at this snapshot, the registry holding 0 real renditions among 4 rendition manifests, the remainder being gate fixtures. Both report UNDEFINED under code NA, not a flattering full mark. Chain availability was checked outside the numerator: of 64 registered objects, 59 recomputed equal. A3, exploratory by registration, now resolves: 88598 of 88598 source characters carry an authorship record across 488 spans, recorded as 0 human-typed, 244 model-generated, 0 quoted-source and 244 script-injected, with 0 back-filled after the fact; the full mark is a consistency check on a ledger the observed party fills about itself. Blind spots as frozen, card by card. A1: "Proves only that the artifact can be mechanically replayed from the source md; it does not prove who wrote the words in the source md, or on what warrant (the latter belongs to A3, exploratory this round); non-semantic byte differences produced by format conversion (PDF/DOCX) do not enter this metric (the object of comparison is limited to the canonical UTF-8 stream before format conversion)." A2: "Governs only the registration completeness of the chain, not the semantic correctness of the chain (a script may be registered in full and still be logically wrong)." A3: "The ledger is self-filled by the observed party and is weak as self-evidence; this round it explicitly supports no confirmatory claim."

Claim-evidence closure decomposes into registration (B1), evidence linkage (B2), source validity (B3) and independent sufficiency audit (B4); the denominator of B1 comes from a deterministic parser, not from the observed party. The register that stood empty when this section was first collected now holds entries, so the four cards report values in place of nulls; each figure is given as a count with the denominator it closes on and the unit stated where it is reported, never composed across units. The pre-committed rule that a structurally present but substantively empty register yields a zero denominator rather than a missing value is unchanged, and is the rule under which the earlier nulls were recorded. Registration (B1) is reported as one frozen ratio, with the register\textquotesingle s own coverage of body sentences given beside it as counts on a different unit and never composed with it. On the unit of candidate claim sentences, the parser\textquotesingle s denominator stands at 398, of which 151 carry a registered claim identifier and 247 carry none. Of the register itself, 167 entries, 108 reach at least one marked body sentence and 59 reach none; each grouping closes on its own denominator. Evidence linkage (B2) stands at 167 of 167 registered claims carrying at least one evidence item, with 0 carrying none, the second figure recorded in its own right so that a full mark is read for what it is. Under the standing clause of Section 3 that value is admitted as a consistency check on the registration procedure and as nothing else. Evidence reuse is disclosed beside the linkage as the card\textquotesingle s anti-gaming rule requires, at 39 evidence-to-source links; the per-evidence citation distribution the same rule requires alongside is not carried on the card, whose note in its place is a fixed string written when the register was still empty and emitted unchanged since, so that distribution is not available at this snapshot. Source status is reported four ways and never merged: 106 hashed and passing, 0 versioned references, 6 pending, 0 failing. Independent sufficiency audit (B4) is reported as adjudicated: 10 of 30 sampled claims were judged to carry sufficient evidence. The sample was drawn by a formula frozen before the audit began, from a registered frame of 167 claims, under a seed generated by someone outside collection whose value enters no report. The caveat frozen on the card requires the NA count alongside: the card carries no separate NA field at this snapshot, so that count is not available here. Anchor defects noticed during adjudication were recorded as correction items and routed to a separate correction process; the frozen rule bars them from altering the verdicts reported here, and we state nothing about what a corrected register would return. Three-path equivalence returned Yes. Blind spots as frozen, card by card. B1: "The criteria of the body parser may themselves fail to recognise difficult claims; the blind-audit sample is small, so the interval on the recall estimate must be reported alongside." B2: "Proves only that an E is attached; it does not prove that the E is relevant, independent or sufficient (sufficiency belongs to B4)." B3: "VERSIONED\_REFERENCE has no local hash, and its validity depends on a version anchor of URL plus publication date; the risk of link rot cannot be captured by this metric, so the two classes must be reported in separate columns." B4: "Human audit is subjective; sample size is constrained by cost; too high a share of NA hollows the rate out, so the NA count must be reported alongside."

Sealing carries two deliberately weak cards. Commit density (D1) was demoted to descriptive before collection because commit granularity is habit, not mechanism; it reports 1 raw commits beside 1 grouped sealing events, with no expected value. Cold-start recovery (D2), one of the cards registered across both cases, reports counts only: a drill count of 1 against 8 checkpoint archives on disk, the drill\textquotesingle s outcome registered as partial rather than rounded up (1), its blocking items carried verbatim on the card, and the by-case breakdown carried in the snapshot. The card freezes the distinction that keeps the count honest: an archive existing is not evidence anyone rehearsed recovery. Blind spots as frozen, card by card. D1: "Commit granularity is a personal habit, not a property of the mechanism; the ICIS case has no repository of its own, so the history of the parent repository must be filtered by path and granularity is even less comparable (see the case roster)." D2: "N is extremely small (single-digit); reporting a rate is forbidden, and statistical inference is forbidden." The case label carried on the D1 card is the one frozen at registration; it denotes the retrospective case, Case 2.

Privacy and caliber gating reports run counts, verdicts and ledger completeness only; intercepted text never enters the observation repository. Prospectively the gate ledger now exists and the three cards resolve: 3 runs over 22 scanned files, 3 blocked, 0 hits falling in a privacy class and the remainder in a caliber class. Override logging (E2) records 3 of 3 blocked runs carrying an adjudication artifact that links back to the run, admissible as a consistency check and not as efficacy. Hit adjudication (E3) registers 234 rulings of false positive against 0 of true positive, from which no precision claim follows. The retrospective ledger yields 55 runs over 163 scanned files, 12 blocked and 1 aborted and purged. Override logging (E2) records 8 of 8 events with all six required elements present at this snapshot, admissible as a consistency check, not as efficacy; it flags 8 rows whose timestamps are prose, not instants. Hit adjudication (E3) is coded MISSING against 12 candidate failing runs, so no false-positive rate is claimed. Blind spots as frozen, card by card. E1: "A high block count may mean the gate is effective and may equally mean input quality is poor; the two cannot be told apart." E2: "Complete record-keeping does not equal correct judgement; the substantive reasonableness of a release rationale is not evaluated by this metric (that belongs to E3 and to human audit)." E3: "There is no independent ground truth, and the false-positive adjudication is itself a human judgement; no precision claim is made."

Human-machine attribution is descriptive by construction, every more-is-better reading removed at design time. Final decision points (F1), a card registered across both cases and reported split by case, stand at 53 adjudicated amendments prospectively and 3 decision artifacts retrospectively; the registered veto-and-redirection count (F2), a card the roster likewise registers across both cases, stands at 6 events, counted only where they point to a discernible artifact difference and broken out by direction on the card rather than composed into a rate. Agent activity (F3) is exploratory and reports raw rows beside de-duplicated task packages, since either alone is gameable: 54 and 54 prospectively, 55 and 40 retrospectively. On the prospective ledger, failed, retried and aborted calls are recorded as not obtainable rather than defaulted to zero, that ledger registering only sessions actually executed; the retrospective ledger does carry an aborted-and-purged count, reported above under privacy gating. With 0 adjudicated adoption verdicts, no adoption rate is computed. Blind spots as frozen, card by card. F1: "Human final adjudication is a registered fact and does not stand for actual cognitive investment; every implication that more is better is removed." F2: "A veto is recorded only on condition that somebody records it; a spoken veto never written to disk is unobservable." F3: "The standing of this card is exploratory, for two reasons: first, the existence of the prospective model-call ledger is decided by probing at collection time and cannot be guaranteed in advance; second, after de-duplication by task package, call count still bears no fixed relation to cognitive contribution."

Reproducibility and cost close the set. Independent rebuild (G1) reports a drill count of 1, performed by a rebuilder outside the production line (1), with the outcome registered as reproduced with deviation (1) and the deviations carried verbatim on the card; N is too small for any rate and none is computed. The presence of an upstream idempotency report is recorded as Yes, but same-environment self-recomputation is excluded from the numerator rather than borrowed as credibility. Cost (G2) is exploratory and asserts no saving; the earlier licence to omit values where obtainable was deleted before collection, so the 2 unobtained fields on this card, another registered across both cases, are enumerated with reasons rather than dropped. Blind spots as frozen, card by card. G1: "N is extremely small; reporting a rate is forbidden, and statistical inference is forbidden; how clean a clean environment is cannot be fully proved." G2: "Wall-clock duration includes interruption and waiting; there is no reliable ledger for token-level cost, and this protocol asserts no cost-saving conclusion (§5 claim boundary)."

MISSING and NA codes remain across the families, and their density is itself a pre-registered reported quantity, not an apology. What remains traces to conditions the protocol provided for in advance: the observation freeze manifest not yet in effect, no body rendition yet registered upstream, and ledgers that did not exist when the cards were frozen. Under the frozen rules such conditions yield nulls, explicit codes and declared degradations, never zeros or full marks. The families support a system claim about what a pre-frozen instrument records, its subjects\textquotesingle{} incompleteness included, within the claim boundary stated in Section 7.

\subsection{Honest Boundaries}\label{honest-boundaries}

We state the gravest limitation first: the observer and the observed are the same team. We built the workflow, ran the projects it governs, wrote the protocol that measures them, authored the collection scripts and adjudicate disputes over their output. The frozen role table separates operator, observer and adjudicator and bars the observer from writing to an observed repository; collection is read-only and hashes upstream files before and after every snapshot, voiding any taken across a change; script hashes are fixed before the first confirmatory event, so retuning one after a result is seen requires a substantive amendment and dual-caliber reporting; blinded audit seeds are generated by someone who does not run collection. None of this dissolves the conflict. In a single-team project the orchestrating role necessarily holds both operator and adjudicator standing, a residual conflict the protocol registers as structural and irreducible rather than isolated. Independent cross-vendor review is a counterweight, not an escape.

There is no control group and could not be one: the same manuscripts cannot be run twice, with and without the discipline. Every claim here is therefore a system claim: a mechanism was in place, ran, produced a recorded verdict, and the verdict propagated. Nothing supports a claim that the discipline made the research better or faster. The frozen text adds three prohibitions: no cost-saving claim, no claim about degree of human control, and no reading of a gate-forced value as evidence of efficacy. The absence of effect estimates is by construction, not by omission.

The case base is small and deliberately unmerged: one case confirmatory and prospective, the second retrospective at an explicitly lower evidential grade, the two never composed. A third and earlier engineering package, the prototype from which the claim--evidence index descends, enters as mechanism lineage only: it supplies no metric data and counts toward no case total. Both cases come from one laboratory, one toolchain and one research programme, so we claim no transfer to teams with other incentives, tools or disciplinary norms. Of the registered rebuild drills, those performed by someone outside the production line number 1; the self-recomputation evidence that exists runs in the same environment and is excluded from that card by rule. Until the observation freeze manifest takes effect, the reported snapshot carries provisional status and pilot/exploratory data class, whatever the observed projects\textquotesingle{} gates return.

Blind spots are not a paragraph but a field, transcribed card by card in Section 6. Replay proves a product can be regenerated from registered sources, not who wrote them or on what warrant -\/- the character-span ledger that would answer that records 0 spans as human-typed, a figure the observed party supplies about itself. Registration completeness says nothing about semantic correctness; an evidence link proves attachment, not relevance or sufficiency; gap counts are incomparable across gates; remediation latency is wall-clock; commit density is personal habit; a high interception count fits an effective gate and poor inputs equally; false-positive adjudication has no ground truth; a registered decision point records that someone decided, not what it cost; an unrecorded override is indistinguishable from one that never happened; drill counts are too small for any rate. An instrument that reports its own missing data still cannot certify its own honesty. One such defect is recorded here rather than quietly repaired: in the post-audit correction period, and outside the two-case structure this paper reports on, a re-adjudication packet built for the human layer of the sufficiency audit passed every machine check that compared it against its frozen source, and was still withdrawn before any human decision was locked or registered, because its cards carried unresolved field placeholders and file locators where resolved values and quoted source were required, leaving no adjudicator able to decide within the card. What those checks established was fidelity to a frozen source, not decision-readiness for the person the packet is rendered for, and the protocol had frozen no check of the second kind; a human-factor acceptance clause was added afterwards, under which no packet is put to an adjudicator until a reviewer holding no prior context has completed a judgement on one card\textquotesingle s face. The human layer was thereupon re-registered, by substantive amendment effective for events not yet occurred, as a process attestation with in-depth review of a subsample: a recomputable machine pass over 167 items of the registered claim set, reported as process checks and not as sufficiency verdicts; in-depth examination of 5 fully rendered cards, purposefully selected as the complete known overlap between the earlier samples rather than drawn at random; and a signed statement in which the same informed adjudicator records having reviewed the machine verification chain and its sealed reports, examined those cards, and taken responsibility for the accurate reporting of the process checks and the item-level decisions. That attestation is not a human re-adjudication of the sampled set, item by item or in aggregate: the sufficiency-audit numerator and denominator reported in Section 6 remain those registered in the first round and nothing in the attestation enters either, the purposefully selected items support no statement about the frame they came from, and the withdrawal is itself carried on the override ledger as one of the 6 registered redirection events -\/- a redirection observable at all because it was written down. What we offer instead is the minimal review package specified in Section 9, with which a third party recomputes every primary metric without us.

\subsection{Discussion}\label{discussion}

A discipline is useful only if parts of it can be adopted separately, so we state a minimum adoptable kit, ordered by design judgement, not measurement: nothing in our data ranks these mechanisms against one another, and the protocol forbids composing them into a score. First, and cheapest, programmatic body injection with a registered build chain --- source, script, and a digest for each --- so that a released document can be replayed in a clean environment and compared byte for byte with the repository copy; it turns "was this number typed by hand" into a question a script answers. Second, append-only retention of every gate attempt, including failures, reruns and manual interventions. Its absence is fatal to measurement, for the reason given by the standing clause in Section 3: without the attempt process there is nothing left to observe but terminal values, and those the protocol refuses to read as efficacy. Third, a claim--evidence index whose denominator is drawn by a deterministic parser under criteria frozen in advance, not by the party whose coverage is measured. Fourth, a machine check where artifacts leave the engineering root, with a mandatory ledger row whenever a person overrides a block. Fifth, a freeze manifest with an amendment ledger, so that revision is visible as revision.

The expensive part of adoption is not the tooling; each element is ordinary scripting. It is the rule that a record, once written, is not rewritten: a failed attempt stays, an override names the releasing role, a superseded anchor is voided in place rather than quietly recut. Role separation is hardest for a small group, and Section 7 states why we have not solved it. What we can defend is narrower: whoever clears an artifact is never whoever produced that artifact, with cross-model review standing in where an independent reviewer cannot be staffed. The signed process attestation registered for the sufficiency audit, described in Section 7, takes its shape from a structure that exists outside research: under Section 302 of the Sarbanes-Oxley Act of 2002, a named officer reviews a disclosure, certifies it, and bears personal responsibility for the controls behind it {[}9{]}. Our borrowing stops at that structure -\/- a named role signs for the accurate reporting of a machine verification chain and of the item-level decisions recorded on the purposefully selected cards, which comprise the complete known overlap between the earlier samples -\/- and asserts no conformity with that statute and no equivalence to it. Placed against the IIA\textquotesingle s Three Lines Model, that signature sits with the roles that operate and control the work rather than with an independent assurance function {[}10{]}: one team supplies operation, control and attestation alike, and the cross-model review named above is not staffed as such a function. The attestation record identifies the person who accepted responsibility for the stated review boundary; it is not a demonstration that the work it covers is good, so we report how this human contribution was registered and how it occurred, and claim nothing beyond this case. We claim nothing about what this costs: our cost card is exploratory, its Case 1 fields are coded MISSING for want of an upstream call ledger and commit-anchored stage boundaries, and the prohibition on cost-saving claims is written into the protocol\textquotesingle s claim boundary.

Four adjacent systems address parts of the same problem. We position against each at the level of design category; no contrast below is a defect claim. DraftMarks (CHI 2026) archives the writing process inside the editor {[}11{]} --- precisely the location where our own instrumentation is weakest: our character-span authorship card is exploratory because its upstream span ledger is registered by the observed party and, even where fully read, cannot establish who actually wrote the text. Our unit of record is the repository and the release gate, which buys coverage of what no editor sees --- data, scripts, ledgers, gate outcomes --- and a record that outlives the editing session, at the price of attribution inside the editing session itself. GitOfThoughts (arXiv 2026) stores reasoning trees in git {[}12{]}; we use the same substrate for a different payload --- not the deliberation but what it produced and what checked it, bound in a manifest a third party recomputes item by item and frozen before observation begins.

PROV-AGENT (IEEE e-Science 2025) carries provenance over agent metadata {[}13{]}. Our sanitized-workspace mechanism records material flow at an engineering boundary instead of an agent\textquotesingle s internal call graph: which files left, which came back, whether each crossing cleared the scan. The two are complements, and the gap runs in our direction: our multi-agent activity card is exploratory precisely because call counts inflate when work is split. Safe-SDL (arXiv 2026) pairs signed logs with autonomy grading {[}14{]}. Two differences follow from construction: our seal\textquotesingle s authority is a hash inventory anyone can recompute, so it does not rest on trusting a signer; and no agent holds release authority, so a human release also leaves a counted row, which makes "who overrode whom" an observable rather than a policy. What distinguishes our work from all four is not a component but a configuration: an end-to-end discipline adoptable with ordinary tooling, evidenced by a protocol frozen before the confirmatory production it observes rather than by a case reconstructed afterwards.

The Introduction advanced, as a design hypothesis rather than a measured result, that the binding constraint on auditability is the marginal cost of recording and curating process, paid by an individual in career time, unfunded, for a benefit accruing largely to others. Two things follow that the Introduction did not say. The first is that machine assistance would move that cost rather than remove it, because writing manifests, maintaining ledgers, binding claims to evidence and re-deriving artifacts from sources is much of the labour it performs at low marginal cost. This remains a hypothesis about where the cost sits, stated so that it can be tested, and not a measurement; our own cost instrumentation is exploratory and partly missing. The second is what the record does establish, which is weaker: the ledgers, the retained failing attempts and a fault-injection suite whose negative cases must fail before a freeze may be generated all exist as by-products of the work, not as documentation written afterwards.

The by-product supplies what the literature lacks. Publication norms record successes; failed attempts and abandoned branches rarely enter the record, and almost never in a form a machine can read. A workflow of this kind emits them as exhaust: in the prospective case, at the provisional snapshot and therefore at pilot/exploratory data class, 6 of 24 retained gate attempts carry at least one recorded gap, counted per attempt and reported by gate. The frozen definition names a gap after the rule that failed; the upstream ledger keeps only per-attempt totals, so that identification remains a property of the protocol rather than of the record we can show. The halt is the sharper instance. Pre-registration is usually defended as a guard against selective reporting after results are seen; this case shows a second function that operates before {[}15{]}. The response to the no-go boolean was written and sealed while its value was unknowable, so when the check fired there was no decision left to take, only an instruction issued by people who did not know which way it would cut. Registered reports institutionalize that structure for a study; our single instance suggests it transfers to a production pipeline, and we claim no more than the instance {[}1{]}. The strongest evidence we can offer for the discipline is the least flattering fact about this paper: it cost us the confirmatory claim we set out to report, and the record of that cost is machine-readable.

\subsection{Resource Availability}\label{resource-availability}

\subsubsection{Lead contact}\label{lead-contact}

Requests for further information and resources should be directed to and will be fulfilled by the lead contact, Yang Zhou (996420470@qq.com).

\subsubsection{Materials availability}\label{materials-availability}

This study did not generate new unique reagents.

\subsubsection{Data and code availability}\label{data-and-code-availability}

\begin{itemize}
\tightlist
\item
  Derived data and metadata will be deposited at Zenodo under no embargo and will be publicly available as of the date of publication. The corresponding repository DOI will be added before acceptance.
\item
  All original code will be deposited at Zenodo under no embargo and will be publicly available as of the date of publication. The corresponding repository DOI will be added before acceptance.
\item
  Any additional information required to reanalyze the data reported in this paper is available from the lead contact upon request.
\end{itemize}

The audit package is this paper\textquotesingle s primary artifact, its inventory fixed before the first confirmatory event: the freeze manifest, every audit snapshot, the collection and metric scripts with their registered digests, the amendment, deviation and attempt ledgers, the case roster, the four mechanism cards, and the pre-registration protocol. Its verifier rebuilds that inventory independently and compares as a set, so omitting one item fails rather than passes; the scripts admit no third-party dependency. A third party can recompute every primary metric from any snapshot without us, and every reported metric value is produced by a read-only runner and injected mechanically, never transcribed. Derived data and metadata will be released under CC BY 4.0, and the scripts under MIT, on the observed project\textquotesingle s terms; neither licence extends to any source document.

What is withheld is withheld by rule. The rights matrix registered upstream has four classes, three of which cover source documents --- officially published and versionable, third-party public with unclear reuse rights, and restricted original --- while the fourth, derived publishable data, comprises the observed project\textquotesingle s own derived outputs and contains no source document. Every source document therefore carries exactly one of the three source-document classes, and material without a registered rights class may not enter any external product. Where redistribution rights are absent or unclear we deposit the URL, retrieval date, digest, metadata and structured extraction rather than the document, never mirror a watermarked original, and release no watermark-circumventing script. Quotation is capped per quotation and per document by a pre-publication check that exits non-zero and cannot be waived by hand. Gate material is deposited as hashes, counts and opaque codes: intercepted text and the detection wordlist never enter the observation repository, review records keep roles and drop persons, and every signer field is institutional.

Sealing is layered and time-bound. Upstream, the blinded item layer opens once that project\textquotesingle s calibration manifest verifies and its access log audits clean, a gate that precedes the blind run; only the result layer is sealed to that project\textquotesingle s release freeze. Every unsealing is recorded append-only with an institutional signer and a witness, and our sampling seed is deposited by digest while its value enters no report. Cards reading sealed or absent upstream ledgers are published with their missing codes and re-collected against the later anchor, never back-filled; corrections are appended as superseding entries and issued as versioned errata, never written over. The halt that followed the confirmatory test changes none of this: sealed results keep their descriptive-not-confirmatory label, the dataset is locked with its termination time, collected data and stated reason, and a takedown procedure for rights holders ships inside the package.

\subsection{Acknowledgments}\label{acknowledgments}

This work was supported by the Key Research and Development Program of the Department of Science and Technology of the Tibet Autonomous Region ("Research on the Science and Technology Development Roadmap of the Tibet Autonomous Region"; application acceptance no. CGZH2025000494).

\subsection{Author contributions}\label{author-contributions}

Y.Z.: Conceptualization; Methodology; Software; Formal analysis; Investigation; Data curation; Writing -- original draft; Writing -- review \& editing; Visualization. C.Y.: Conceptualization; Resources; Supervision; Project administration; Funding acquisition; Writing -- review \& editing.

\subsection{Declaration of interests}\label{declaration-of-interests}

Chengqun Yu serves as a project review expert for the Department of Science and Technology of the Tibet Autonomous Region, but was not involved in the review or award of the project supporting this work. Yang Zhou declares no competing interests.

\subsection{Declaration of generative AI and AI-assisted technologies in the writing process}\label{declaration-of-generative-ai-and-ai-assisted-technologies-in-the-writing-process}

During the preparation of this work the authors used Anthropic Claude (claude-opus-5 and claude-fable-5), OpenAI Codex, and Moonshot AI Kimi K3 in order to improve the readability and language of the manuscript. After using these tools and services, the authors reviewed and edited the content as needed and take full responsibility for the content of the published article.

Model participation in drafting, cross-model criticism, consistency checking, and the development of auditable assembly code forms part of the research process rather than the writing process --- the object of study of this paper --- and is reported in the mechanism and protocol sections.

No generative AI image tool created or modified any research or data image in this manuscript, and no such tool was used to adjust brightness, contrast or colour balance; AI assistance was used to help write deterministic figure-building code, and the figures themselves were produced from registered sources by those scripts. Generative AI and AI-assisted technologies are not listed as authors of this work and are not cited as authors.

\subsection{References}\label{references}

{1.} Chambers, C.D., and Tzavella, L. (2022). The past, present and future of Registered Reports. \emph{Nature Human Behaviour} 6, 29--42. \url{https://doi.org/10.1038/s41562-021-01193-7}

{2.} DeAngelis, C.D., Drazen, J.M., Frizelle, F.A., Haug, C., Hoey, J., Horton, R., Kotzin, S., Laine, C., Marusic, A., Overbeke, A.J.P.M., Schroeder, T.V., Sox, H.C., and Van Der Weyden, M.B. (2004). Clinical trial registration: a statement from the International Committee of Medical Journal Editors. \emph{JAMA} 292, 1363--1364. \url{https://doi.org/10.1001/jama.292.11.1363}

{3.} De Angelis, C.D., Drazen, J.M., Frizelle, F.A., Haug, C., Hoey, J., Horton, R., Kotzin, S., Laine, C., Marusic, A., Overbeke, A.J.P.M., Schroeder, T.V., Sox, H.C., and Van Der Weyden, M.B. (2005). Is This Clinical Trial Fully Registered? --- A Statement from the International Committee of Medical Journal Editors. \emph{New England Journal of Medicine} 352, 2436--2438. \url{https://doi.org/10.1056/NEJMe058127}

{4.} Nuijten, M.B., Hartgerink, C.H.J., van Assen, M.A.L.M., Epskamp, S., and Wicherts, J.M. (2016). The prevalence of statistical reporting errors in psychology (1985--2013). \emph{Behavior Research Methods} 48, 1205--1226. \url{https://doi.org/10.3758/s13428-015-0664-2}

{5.} Groth, P.T., Gibson, A., and Velterop, J. (2010). The anatomy of a nanopublication. \emph{Information Services and Use} 30, 51--56. \url{https://doi.org/10.3233/ISU-2010-0613}

{6.} Kuhn, T., Meroño-Peñuela, A., Malic, A., Poelen, J.H., Hurlbert, A.H., Centeno Ortiz, E., Furlong, L.I., Queralt-Rosinach, N., Chichester, C., Banda, J.M., Willighagen, E., Ehrhart, F., Evelo, C., Malas, T.B., and Dumontier, M. (2018). Nanopublications: A Growing Resource of Provenance-Centric Scientific Linked Data. In \emph{2018 IEEE 14th International Conference on e-Science (e-Science)} (pp. 83--92). IEEE. \url{https://doi.org/10.1109/eScience.2018.00024}

{7.} Nicholson, J.M., Mordaunt, M., Lopez, P., Uppala, A., Rosati, D., Rodrigues, N.P., Grabitz, P., and Rife, S.C. (2021). scite: A smart citation index that displays the context of citations and classifies their intent using deep learning. \emph{Quantitative Science Studies} 2, 882--898. \url{https://doi.org/10.1162/qss_a_00146}

{8.} Shotton, D. (2010). CiTO, the Citation Typing Ontology. \emph{Journal of Biomedical Semantics} 1, S6. \url{https://doi.org/10.1186/2041-1480-1-S1-S6}

{9.} United States Congress. (2002). Sarbanes--Oxley Act of 2002. \url{https://www.govinfo.gov/link/plaw/107/public/204}

{10.} The Institute of Internal Auditors. (2026). Three Lines Model: Assurance and Advice in Support of Effective Governance. \url{https://www.theiia.org/globalassets/site/resources/statements-of-position/tlm_assurance_advice_support_effective_gov_en.pdf}

{11.} Siddiqui, M.N., Nasseri, N., Coscia, A.J., Pea, R., and Subramonyam, H. (2026). DraftMarks: Enhancing Transparency in Human-AI Co-Writing Through Interactive Skeuomorphic Process Traces. In \emph{Proceedings of the 2026 CHI Conference on Human Factors in Computing Systems (CHI \textquotesingle26)} (pp. 1--22). ACM. \url{https://doi.org/10.1145/3772318.3791109}

{12.} Shekar, P.C., H S, A., and Krishnan, A. (2026). GitOfThoughts: Version-Controlled Reasoning and Agent Memory You Can Replay, Diff, and Merge. \emph{arXiv}. \url{https://doi.org/10.48550/arXiv.2606.14470}

{13.} Souza, R., Gueroudji, A., DeWitt, S., Rosendo, D., Ghosal, T., Ross, R., Balaprakash, P., and Ferreira da Silva, R. (2025). PROV-AGENT: Unified Provenance for Tracking AI Agent Interactions in Agentic Workflows. In \emph{Proceedings of the 21st IEEE International Conference on e-Science (e-Science)} (pp. 467--473). IEEE. \url{https://doi.org/10.1109/eScience65000.2025.00093}

{14.} Zhang, Z., Que, H., Chang, J., Zhang, X., Wei, H., and Zhu, T. (2026). Safe-SDL: Establishing Safety Boundaries and Control Mechanisms for AI-Driven Self-Driving Laboratories. \emph{arXiv}. \url{https://doi.org/10.48550/arXiv.2602.15061}

{15.} Nosek, B.A., Ebersole, C.R., DeHaven, A.C., and Mellor, D.T. (2018). The preregistration revolution. \emph{Proceedings of the National Academy of Sciences} 115, 2600--2606. \url{https://doi.org/10.1073/pnas.1708274114}

\end{document}